\documentclass[10pt,english,british,tightenlines,eqsecnum,
floats,aps,amsmath,amssymb,nofootinbib,superscriptaddress,prd,showpacs,showkeys]
              {revtex4}
\usepackage[latin9]{inputenc}
\usepackage{color}
\usepackage{babel}
\usepackage{amsmath}
\usepackage{amssymb}
\usepackage{graphicx}
\usepackage{esint}
\usepackage[unicode=true,pdfusetitle,
 bookmarks=true,bookmarksnumbered=false,bookmarksopen=false,
 breaklinks=false,pdfborder={0 0 1},backref=false,colorlinks=false]
 {hyperref}

\makeatletter
\@ifundefined{textcolor}{}
{%
 \definecolor{BLACK}{gray}{0}
 \definecolor{WHITE}{gray}{1}
 \definecolor{RED}{rgb}{1,0,0}
 \definecolor{GREEN}{rgb}{0,1,0}
 \definecolor{BLUE}{rgb}{0,0,1}
 \definecolor{CYAN}{cmyk}{1,0,0,0}
 \definecolor{MAGENTA}{cmyk}{0,1,0,0}
 \definecolor{YELLOW}{cmyk}{0,0,1,0}
}

\usepackage{babel}

\usepackage{euscript}\usepackage{epsfig}

\usepackage{amsfonts}

\usepackage{fancyhdr}

\makeatother

\begin{document}

\title{Topological Diagnostics of CP Degeneracies in Long-Baseline Neutrino Oscillations with Non-Standard Interactions via Pancharatnam Open-Path Phases}

\author{N. Razzaghi}
\email{n.razzaghi@iau.ac.ir}

\affiliation{Department of Physics, QA.C., Islamic Azad
University, Qazvin, Iran}

\begin{abstract}
We investigate the topological signatures of neutral-current
Non-Standard Interactions (NSI) in long-baseline neutrino
oscillations by extending the open-path Pancharatnam geometric
phase framework. Incorporating off-diagonal vector couplings
($\varepsilon_{e\tau} = |\varepsilon_{e\tau}|e^{i\phi_{e\tau}}$)
into the discrete-layer matter Hamiltonian across the DUNE
baseline ($L = 1300\text{ km}$), we analyze the state evolution
for both neutrino ($\nu_\mu \to \nu_e$) and antineutrino
($\bar{\nu}_\mu \to \bar{\nu}_e$) appearance channels over the
broadband energy spectrum $E_\nu \in [0.5, 8.0]\text{ GeV}$. Our
analysis demonstrates that complex NSI phases imprint distinct,
energy-dependent spectral dispersions onto the open-path geometric
holonomies $\Phi_g^{(n)}$ and their $\mathcal{CP}$-conjugate
counterparts. By clarifying that these geometric phases manifest
through gauge-invariant interference terms $\Delta\Gamma_{nm}$
rather than asymptotic single-channel measurements, we construct
the $\mathcal{CP}$-violating geometric phase differences
$\Delta\Phi_g^{\mathrm{CP}}$. We show that the interplay between
these geometric dispersions and the matter-potential sign reversal
under $\mathcal{CP}$-conjugation forces the neutrino and
antineutrino degeneracy bands to intersect at non-parallel angles
in the two-dimensional $(\delta_{\mathrm{CP}}, \phi_{e\tau})$
parameter space. This non-trivial topological intersection
isolates a unique physical solution, providing a robust,
independent diagnostic capable of breaking persistent parameter
degeneracies even under realistic calorimetric energy smearing.
\end{abstract}

\keywords{Neutrino oscillations; Open-path geometric phase;
Pancharatnam phase; Non-Standard Interactions (NSI); Parameter
degeneracies; Long-baseline experiments (DUNE).
 }

\date{Nov 2, 2026}
\maketitle

\section{Introduction}
\label{sec:introduction}

The discovery of neutrino oscillations has established beyond
doubt that neutrinos possess non-zero masses and mix under
charged-current weak interactions, offering a pristine laboratory
signature of physics beyond the Standard Model
(SM)~\cite{Esteban:2024hjw}. The primary objective of
next-generation long-baseline oscillation facilities, most notably
the Deep Underground Neutrino Experiment
(DUNE)~\cite{Abi:2020evt}, is the high-precision determination of
leptonic Dirac $CP$ violation parameterized by the phase
$\delta_{\text{CP}}$, alongside the definitive resolution of the
neutrino mass ordering.

However, neutrino propagation through terrestrial matter
introduces effective forward coherent scattering potentials that
can be altered by neutral-current Non-Standard Interactions
(NSI)~\cite{Wolfenstein:1978ue, Farzan:2017xzy}. When
parameterized via effective dimension-6 vector operators, these
interactions yield off-diagonal complex matrix elements
$\varepsilon_{\alpha\beta} =
|\varepsilon_{\alpha\beta}|e^{i\phi_{\alpha\beta}}$ in the flavor
Hamiltonian. These auxiliary sources of $CP$ violation can
severely mimic, mask, or counteract the standard leptonic phase
$\delta_{\text{CP}}$~\cite{Coloma:2015kiu, Deepthi:2017gxd}. This
interplay gives rise to acute parameter degeneracies in the
appearance probabilities $P(\nu_\mu \to \nu_e)$ and
$\bar{P}(\bar{\nu}_\mu \to \bar{\nu}_e)$. In standard experimental
analyses, these ambiguities typically manifest as continuous,
overlapping contours in multi-parameter confidence regions,
complicating the unambiguous extraction of genuine Dirac $CP$
violation from monochromatic or narrow-band event rates.

Addressing these degeneracies necessitates looking beyond scalar
transition probabilities into the underlying quantum phase
geometry governing the state evolution. While geometric phases
have been extensively investigated in condensed matter and quantum
optics, their application to neutrino oscillations has
historically centered on the cyclic, adiabatic Berry phase
framework or simplified two-flavor paradigms. These idealized
approximations, however, break down across non-adiabatic MSW
resonances, multi-layer density boundaries, or open-path
evolutionary trajectories. To address these limitations, the
kinematic, open-path Pancharatnam--Mukunda--Simon formulation was
recently established as a gauge-invariant theoretical diagnostic
for three-flavor neutrino propagation across layered media within
standard matter~\cite{Razzaghi:2026ptep}.

In this work, we systematically generalize the open-path geometric
phase formalism to accommodate neutral-current vector NSI within a
rigorous discrete-layer transfer architecture. Unlike traditional
approaches that treat parameter degeneracies at the level of
integrated transition rates, our approach traces how new physics
deforms the underlying state holonomy and eigenstate interference
channels. We focus our phenomenological study on the off-diagonal
coupling $\varepsilon_{e\tau}$, which provides the dominant
interference contribution in the $\nu_\mu \to \nu_e$ appearance
channel. Although the discrete-layer framework is designed to
capture non-adiabatic boundary effects across arbitrary media, we
demonstrate that for the specific continental crust trajectory of
DUNE ($L = 1300$\,km), the adiabatic condition is exceptionally
well preserved, allowing an accurate reduction to a single-layer
three-branch representation without loss of topological rigor.

By analyzing the continuous spectral dispersion across broad
neutrino and antineutrino energy windows, we demonstrate that the
open-path Pancharatnam holonomies $\Phi_g^{(n)}(E_\nu)$ possess
unique energy gradients that cannot be replicated by vacuum Dirac
phase shifts. Furthermore, the sign flip of the matter potential
under $\mathcal{CP}$-conjugation breaks the symmetry between the
neutrino and antineutrino geometric paths. When translated into
the two-dimensional $(\delta_{\mathrm{CP}}, \phi_{e\tau})$
parameter space, these distinct holonomy gradients force the
neutrino and antineutrino degeneracy bands to intersect at
non-parallel angles, isolating a unique physical coordinate and
explicitly resolving the continuous parameter degeneracy under
realistic experimental configurations.

The remainder of this paper is organized as follows. In
Sec.~\ref{sec:formalism}, we establish the open-path geometric
phase formalism for layered media in the presence of general
vector NSI and derive the gauge-invariant probability
decomposition. Section~\ref{sec:numerical_analysis} presents the
numerical evaluations under benchmark DUNE conditions, tracking
the branch holonomies, nodal singularities, and dispersion
profiles, followed by the construction of the two-dimensional
$(\delta_{\mathrm{CP}}, \phi_{e\tau})$ degeneracy contours. In
Sec.~\ref{sec:discussion}, we discuss the experimental
implications of the geometric phase on reconstructed detector
event rates, the impact of finite liquid argon calorimetric energy
resolution, and the robustness of the degeneracy resolution.
Finally, Sec.~\ref{sec:conclusions} summarizes our primary
conclusions and outlines future directions.

\section{Theoretical Formalism: NSI and Open-Path Geometric Phases}
\label{sec:formalism}

\subsection{Neutrino Evolution with Vector NSI}
The spatial evolution of an ultrarelativistic three-flavor
neutrino state vector $|\nu(x)\rangle$ propagating through matter
along the spatial coordinate $x$ is governed by the
Schrödinger-like evolution equation (in natural units $\hbar = c =
1$):
\begin{equation}
    i \frac{d}{dx} |\nu(x)\rangle = \mathcal{H}(x) |\nu(x)\rangle,
    \label{eq:schrodinger}
\end{equation}
where the total effective Hamiltonian in the flavor basis
$\{\nu_e, \nu_\mu, \nu_\tau\}$ decomposes into vacuum and matter
contributions:
\begin{equation}
    \mathcal{H}(x) = \mathcal{H}_{\text{vac}} + \mathcal{H}_{\text{matter}}(x).
    \label{eq:hamiltonian_total}
\end{equation}
The vacuum Hamiltonian is parameterized in terms of the
mass-squared differences and the unitary
Pontecorvo--Maki--Nakagawa--Sakata (PMNS) mixing matrix:
\begin{equation}
    \mathcal{H}_{\text{vac}} = \frac{1}{2E_\nu} U_{\text{PMNS}} \operatorname{diag}\left(0, \Delta m_{21}^2, \Delta m_{31}^2\right) U_{\text{PMNS}}^\dagger,
\end{equation}
where $U_{\text{PMNS}}$ is parameterized by the three mixing
angles $(\theta_{12}, \theta_{13}, \theta_{23})$ and the Dirac
CP-violating phase $\delta_{\text{CP}}$.

The coherent forward scattering through terrestrial matter
introduces contributions from both the standard charged-current
(CC) weak potential and neutral-current vector Non-Standard
Interactions (NSI) mediated by heavy new-physics degrees of
freedom. The matter-induced contribution to the flavor Hamiltonian
is partitioned as
\begin{equation}
    \mathcal{H}_{\text{matter}}(x) = \mathcal{V}_{\mathrm{CC}}(x) + \mathcal{V}_{\mathrm{NSI}}(x).
    \label{eq:matter_partition}
\end{equation}
The standard charged-current potential is given by
\begin{equation}
    \mathcal{V}_{\mathrm{CC}}(x) = \pm V_{\mathrm{CC}}(x) \operatorname{diag}(1, 0, 0), \qquad V_{\mathrm{CC}}(x) = \sqrt{2} G_F N_e(x),
    \label{eq:V_CC}
\end{equation}
where $G_F$ is the Fermi coupling constant, $N_e(x)$ is the
ambient electron number density, and the upper (lower) sign
corresponds to neutrino (antineutrino) propagation. The standard
neutral-current (NC) forward scattering is flavor-universal,
contributing identically across all active states via a term
proportional to the $3\times 3$ identity matrix; because this
generates only an unobservable overall dynamical phase, it is
conventionally subtracted from $\mathcal{H}_{\text{matter}}$.

At energies well below the electroweak scale, vector
neutral-current NSI can be described within an effective field
theory framework by the dimension-6 four-fermion Lagrangian:
\begin{equation}
    \mathcal{L}_{\mathrm{NC}\text{-}\mathrm{NSI}} = -2\sqrt{2} G_F \sum_{f=e,u,d} \varepsilon_{\alpha\beta}^{f,V}
    \left(\bar{\nu}_\alpha \gamma^\mu P_L \nu_\beta\right) \left(\bar{f} \gamma_\mu f\right),
    \label{eq:nsi_lagrangian}
\end{equation}
where $P_L = (1-\gamma_5)/2$ is the chiral projection operator,
and $\varepsilon_{\alpha\beta}^{f,V} =
\varepsilon_{\alpha\beta}^{f,L} + \varepsilon_{\alpha\beta}^{f,R}$
denotes the vector coupling to the matter fermion $f \in \{e, u,
d\}$. Coherent forward scattering in unpolarized neutral Earth
matter then generates the effective NSI potential matrix:
\begin{equation}
    \mathcal{V}_{\mathrm{NSI}}(x) = \pm V_{\mathrm{CC}}(x)
    \begin{pmatrix}
        \varepsilon_{ee}(x) & \varepsilon_{e\mu}(x) & \varepsilon_{e\tau}(x) \\
        \varepsilon_{e\mu}^*(x) & \varepsilon_{\mu\mu}(x) & \varepsilon_{\mu\tau}(x) \\
        \varepsilon_{e\tau}^*(x) & \varepsilon_{\mu\tau}^*(x) & \varepsilon_{\tau\tau}(x)
    \end{pmatrix},
    \label{eq:nsi_hamiltonian}
\end{equation}
where the effective dimensionless couplings
$\varepsilon_{\alpha\beta}(x)$ are normalized to the ambient
electron density:
\begin{equation}
    \varepsilon_{\alpha\beta}(x) \equiv \sum_{f=e,u,d} \frac{N_f(x)}{N_e(x)} \varepsilon_{\alpha\beta}^{f,V} = \varepsilon_{\alpha\beta}^{e,V}
    + \left(1 + \frac{N_n(x)}{N_e(x)}\right) \varepsilon_{\alpha\beta}^{u,V} + \left(2 + \frac{N_n(x)}{N_e(x)}\right) \varepsilon_{\alpha\beta}^{d,V}.
    \label{eq:nsi_effective_couplings}
\end{equation}
Here, $N_u = 2N_p + N_n$ and $N_d = N_p + 2N_n$ have been
expressed in terms of the proton and neutron densities, satisfying
$N_p = N_e$ by charge neutrality. For the continental Earth crust,
the ratio is governed by the electron fraction $Y_e =
N_e/(N_p+N_n) = 0.494$, yielding $N_n/N_e = (1-Y_e)/Y_e \approx
1.024$. Under this nearly isoscalar condition, the effective
couplings $\varepsilon_{\alpha\beta}$ remain uniform along the
baseline. Hermiticity of $\mathcal{H}(x)$ requires
$\varepsilon_{\alpha\beta} = \varepsilon_{\beta\alpha}^*$, meaning
the diagonal terms are purely real while the off-diagonal entries
are generically complex, parameterized as
$\varepsilon_{\alpha\beta} = |\varepsilon_{\alpha\beta}|
e^{i\phi_{\alpha\beta}}$. For antineutrino modes, in addition to
the global sign reversal in Eq.~\eqref{eq:nsi_hamiltonian}, the
vacuum mixing matrix undergoes complex conjugation
($U_{\text{PMNS}} \to U_{\text{PMNS}}^*$, mapping
$\delta_{\text{CP}} \to -\delta_{\text{CP}}$), and the NSI
couplings map as $\varepsilon_{\alpha\beta} \to
\varepsilon_{\alpha\beta}^* = |\varepsilon_{\alpha\beta}|
e^{-i\phi_{\alpha\beta}}$ (implying $\phi_{\alpha\beta} \to
-\phi_{\alpha\beta}$).

In this work, we focus on the off-diagonal parameter
$\varepsilon_{e\tau} = |\varepsilon_{e\tau}| e^{i\phi_{e\tau}}$,
setting all other non-standard entries to zero to directly isolate
its interference geometry and topological degeneracies with the
standard Dirac phase $\delta_{\text{CP}}$.

The transition from neutrino to antineutrino propagation is
strictly governed by the CP-conjugation transformation
$\mathcal{CP}$, which acts on the effective Hamiltonian as
$\bar{\mathcal{H}}(x) = \mathcal{H}^*(x)$. Formally, this
operation enforces the parameter mapping $\delta_{\text{CP}} \to
-\delta_{\text{CP}}$ and $\phi_{\alpha\beta} \to
-\phi_{\alpha\beta}$, accompanied by the sign reversal of the
matter potential $V_{\mathrm{CC}} \to -V_{\mathrm{CC}}$. This
transformation ensures that the Hamiltonian maintains its
Hermitian structure and consistently reflects the CP-invariance
properties of the underlying interaction Lagrangian. Consequently,
the appearance probabilities for antineutrinos,
$\bar{P}(\bar{\nu}_\mu \to \bar{\nu}_e)$, are calculated using the
CP-conjugated Hamiltonian, thereby capturing the genuine chiral
asymmetry induced by the combined action of the matter potential
and the complex NSI phase.

\subsection{Open-Path Pancharatnam Geometric Phase in Non-Uniform Media}
For a medium discretized into $N$ piecewise constant layers (or an
effective profile along the Earth's chord), the local Hamiltonian
in layer $k$ admits the spectral decomposition:
\begin{equation}
    \mathcal{H}_k = W_k D_k W_k^\dagger, \quad D_k = \operatorname{diag}\left(\lambda_1^{(k)}, \lambda_2^{(k)}, \lambda_3^{(k)}\right),
\end{equation}
where $W_k$ represents the unitary matrix diagonalizing the
effective Hamiltonian in layer $k$, and $\lambda_n^{(k)}$ are the
corresponding local matter energy eigenvalues. The state evolution
across layer boundaries is governed by the interface transfer
matrices:
\begin{equation}
    S_{k+1,k} = W_{k+1}^\dagger W_k.
\end{equation}
Let the trajectory be divided into $N$ layers of lengths $\ell_k$,
with $\sum_{k=1}^{N}\ell_k=L$. In layer $k$, the evolution
operator is
\begin{equation}
    \mathcal{U}_k
    = W_k e^{-iD_k\ell_k}W_k^\dagger .
    \label{eq:layer_evolution}
\end{equation}
For propagation from layer $1$ to layer $N$, the total evolution
operator is therefore
\begin{equation}
    \mathcal{U}(L,0)
    = \mathcal{U}_N\mathcal{U}_{N-1}\cdots\mathcal{U}_1 .
    \label{eq:total_layer_evolution}
\end{equation}
The matrix element of the interface transfer matrix is the overlap
between the corresponding eigenstates in adjacent layers,
\begin{equation}
    \bigl[S_{k+1,k}\bigr]_{n_{k+1}n_k}
    = \langle n_{k+1}^{(k+1)}|n_k^{(k)}\rangle .
\end{equation}
These overlaps arise from inserting complete matter-eigenstate
bases at the layer boundaries; they do not represent additional
interactions localized at the interfaces.

A propagation chain is specified by the sequence
\begin{equation}
    \mathcal{C}=(n_1,n_2,\ldots,n_N),
    \qquad n_k\in\{1,2,3\},
\end{equation}
where $n_k$ labels the matter eigenstate in layer $k$. Inserting
the eigenstate resolution of the identity at each interface gives
the exact transition amplitude for the piecewise-constant profile:
\begin{align}
    \mathcal{A}_{\alpha\beta}(E_\nu,L)
    &\equiv
    \langle \nu_\beta|\mathcal{U}(L,0)|\nu_\alpha\rangle
    \nonumber\\
    &= \sum_{n_1,\ldots,n_N=1}^{3}
    \langle \nu_\beta|n_N^{(N)}\rangle
    \left[
        \prod_{k=1}^{N-1}
        \langle n_{k+1}^{(k+1)}|n_k^{(k)}\rangle
    \right]
    \exp\!\left[-i\sum_{k=1}^{N}
        \lambda_{n_k}^{(k)}\ell_k\right]
    \langle n_1^{(1)}|\nu_\alpha\rangle
    \nonumber\\
    &\equiv \sum_{\mathcal{C}}\mathcal{G}_{\mathcal{C}} .
    \label{eq:amplitude_chain_sum}
\end{align}
It is useful to separate the endpoint and interface overlaps from
the dynamical phase. For each chain, define
\begin{equation}
    \mathcal{Z}_{\mathcal{C}}
    \equiv
    \langle \nu_\beta|n_N^{(N)}\rangle
    \left[
        \prod_{k=1}^{N-1}
        \langle n_{k+1}^{(k+1)}|n_k^{(k)}\rangle
    \right]
    \langle n_1^{(1)}|\nu_\alpha\rangle ,
    \label{eq:chain_overlap_factor}
\end{equation}
and
\begin{equation}
    \Phi_{\mathrm{dyn}}^{\mathcal{C}}
    \equiv
    \sum_{k=1}^{N}\lambda_{n_k}^{(k)}\ell_k,
    \qquad
    \Phi_g^{\mathcal{C}}
    \equiv \arg\mathcal{Z}_{\mathcal{C}}
    \pmod{2\pi}.
    \label{eq:chain_phases}
\end{equation}
The contribution of a chain to the amplitude is then
\begin{equation}
    \mathcal{G}_{\mathcal{C}}
    = \mathcal{Z}_{\mathcal{C}}
      e^{-i\Phi_{\mathrm{dyn}}^{\mathcal{C}}}
    = |\mathcal{Z}_{\mathcal{C}}|
      e^{i(\Phi_g^{\mathcal{C}}-\Phi_{\mathrm{dyn}}^{\mathcal{C}})},
    \qquad
    |\mathcal{G}_{\mathcal{C}}|=|\mathcal{Z}_{\mathcal{C}}|.
    \label{eq:chain_amplitude}
\end{equation}
Thus, the magnitude assigned to a chain is explicitly
\begin{equation}
    |\mathcal{G}_{\mathcal{C}}|
    =
    |\langle \nu_\beta|n_N^{(N)}\rangle|
    \left[
        \prod_{k=1}^{N-1}
        |\langle n_{k+1}^{(k+1)}|n_k^{(k)}\rangle|
    \right]
    |\langle n_1^{(1)}|\nu_\alpha\rangle|.
    \label{eq:chain_weight}
\end{equation}

The phase $\Phi_g^{\mathcal{C}}$ is invariant under independent
rephasings of the local eigenvectors,
\begin{equation}
    |n_k^{(k)}\rangle
    \longrightarrow
    e^{i\chi_{n_k}^{(k)}}|n_k^{(k)}\rangle,
\end{equation}
because the phases acquired by the endpoint and interface overlaps
cancel in their product $\mathcal{Z}_{\mathcal{C}}$. Accordingly,
the probability can be written as an interference sum over pairs
of chains:
\begin{equation}
    P(\nu_\alpha\to\nu_\beta)
    =
    \sum_{\mathcal{C}}|\mathcal{G}_{\mathcal{C}}|^2
    +2\sum_{\mathcal{C}<\mathcal{C}'}
    |\mathcal{G}_{\mathcal{C}}|
    |\mathcal{G}_{\mathcal{C}'}|
    \cos\!\left(
        \Delta\Phi_{\mathrm{dyn}}^{\mathcal{C}\mathcal{C}'}
        +\Delta\Gamma_{\mathcal{C}\mathcal{C}'}
    \right),
    \label{eq:prob_decomp}
\end{equation}
where
\begin{equation}
    \Delta\Phi_{\mathrm{dyn}}^{\mathcal{C}\mathcal{C}'}
    \equiv
    \Phi_{\mathrm{dyn}}^{\mathcal{C}'}
    -\Phi_{\mathrm{dyn}}^{\mathcal{C}},
    \qquad
    \Delta\Gamma_{\mathcal{C}\mathcal{C}'}
    \equiv
    \Phi_g^{\mathcal{C}}-\Phi_g^{\mathcal{C}'}
    \pmod{2\pi}.
    \label{eq:relative_chain_phases}
\end{equation}
Here, $\mathcal{C}<\mathcal{C}'$ denotes an arbitrary fixed
ordering of distinct chains, so that each unordered pair is
counted once. The chain expansion is an exact rewriting of the
layered evolution operator; in numerical calculations, the product
of the $3\times3$ layer operators in
Eq.~\eqref{eq:total_layer_evolution} can be evaluated directly
without explicitly enumerating all $3^N$ chains.

For a uniform medium, there is a single layer and each chain is
labelled by one eigenstate $n$. The amplitude reduces to
\begin{equation}
    \mathcal{A}_{\alpha\beta}(E_\nu,L)
    =
    \sum_{n=1}^{3}
    \langle \nu_\beta|n\rangle
    \langle n|\nu_\alpha\rangle
    e^{-i\lambda_n L},
    \label{eq:uniform_amplitude}
\end{equation}

and the corresponding branch-dependent geometric phase is
\begin{equation}
    \Phi_g^{(n)}
    =
    \arg\!\left[
        \langle \nu_\beta|n\rangle
        \langle n|\nu_\alpha\rangle
    \right]
    \pmod{2\pi}.
    \label{eq:branch_phase}
\end{equation}

For a single-layer uniform baseline, the relative open-path
geometric phase between branches $n$ and $m$ is simply given by
$\Delta\Gamma_{nm} \equiv \Phi_g^{(n)} - \Phi_g^{(m)}
\pmod{2\pi}$.

\paragraph{Phase branches and nodal points.}
The phase convention used below distinguishes the principal phase
from its continuously unwrapped representative. For a given
propagation branch, let $z_n(E_\nu)$ denote the corresponding
complex transition-amplitude contribution, including its kinematic
prefactor, dynamical phase, and matter-eigenstate overlaps. The
principal phase is defined as
\begin{equation}
\theta_n^{\mathrm{pr}}(E_\nu) =
\operatorname{Arg}\!\left[z_n(E_\nu)\right], \qquad
\theta_n^{\mathrm{pr}} \in (-\pi,\pi].
\label{eq:principal_branch_phase}
\end{equation}
The same convention is applied to the total transition amplitude
$z_{\alpha\beta}(E_\nu)=\mathcal{A}_{\alpha\beta}(E_\nu)$ and to
the relative holonomies constructed from the corresponding branch
amplitudes.

For a discretized energy grid $E_i$, the unwrapped phase is
obtained by adding the smallest principal phase increment between
neighboring points,
\begin{equation}
\widetilde{\theta}_n(E_i) = \widetilde{\theta}_n(E_{i-1}) +
\operatorname{PV}_{(-\pi,\pi]} \left[ \theta_n^{\mathrm{pr}}(E_i)
- \theta_n^{\mathrm{pr}}(E_{i-1}) \right],
\label{eq:phase_unwrapping}
\end{equation}
where $\operatorname{PV}_{(-\pi,\pi]}$ denotes reduction to the
interval $(-\pi,\pi]$. This prescription follows a phase branch
continuously as long as the corresponding complex amplitude does
not vanish.

At a nodal point, defined by
\begin{equation}
z_n(E_\star)=0, \label{eq:nodal_condition}
\end{equation}
the Pancharatnam phase of that contribution is undefined.
Consequently, a phase curve cannot be continued through $E_\star$
without specifying a separate branch convention. In a numerical
calculation, points for which $|z_n(E_\nu)|$ falls below the
numerical nodal-resolution threshold are therefore identified as
nodal points and are not interpreted as physical discontinuities
of the propagation amplitude.

For a simple zero, the local behavior may be written as
\begin{equation}
z_n(E_\nu) \simeq c_n(E_\nu-E_\star), \qquad c_n\neq 0.
\label{eq:local_nodal_behavior}
\end{equation}
When the trajectory in the complex-amplitude plane crosses the
origin, the principal phase generally changes by $\pi$ across the
zero. The apparent jump is a consequence of the chosen branch cut
of $\operatorname{Arg}$; it does not represent a finite phase
accumulated at the nodal point itself. If the trajectory passes
close to, but not exactly through, the origin, the phase instead
undergoes a rapid but continuous rotation, whose detailed
magnitude depends on the local path in the complex plane.

The geometric phase is treated in the same manner. Its principal
representative is
\begin{equation}
\gamma_g^{\mathrm{pr}}(E_\nu) = \operatorname{PV}_{(-\pi,\pi]}
\left[ \Phi_{\mathrm{tot}}(E_\nu) - \Phi_{\mathrm{dyn}}(E_\nu)
\right], \label{eq:principal_geometric_phase}
\end{equation}
whereas the continuously followed quantity is denoted by
$\widetilde{\gamma}_g(E_\nu)$ and is obtained by the unwrapping
prescription in Eq.~\eqref{eq:phase_unwrapping}. Thus, a phase
slip by $\pi$ or $2\pi$ in a principal-value plot is not by itself
evidence for a discontinuous physical evolution. The physically
relevant distinction is whether the complex amplitude crosses the
origin or merely passes close to it.

A vortex interpretation requires a two-dimensional parameter
space, such as $(E_\nu,\delta_{\mathrm{CP}})$ or
$(E_\nu,\phi_{e\tau})$. In that case, the winding number around an
isolated zero of the complex amplitude is
\begin{equation}
\nu_{\mathrm{vortex}} = \frac{1}{2\pi} \oint_{\mathcal{C}} \nabla
\operatorname{Arg}\!\left[z_n\right] \cdot d\boldsymbol{\lambda},
\label{eq:vortex_winding_number}
\end{equation}
where $\boldsymbol{\lambda}$ denotes the chosen two-dimensional
parameter vector and $\mathcal{C}$ is a closed contour that
encloses the zero but no other singularity. In the one-dimensional
energy scans considered in Fig.~\ref{fig:nsi_pancharatnam}, we
refer more conservatively to nodal points and phase singularities
rather than assigning a vortex charge directly.

It is important to emphasize that the phase of an isolated
amplitude contribution is not separately observable at an exact
nodal point, because the corresponding contribution vanishes
there. The observable quantity remains the total transition
probability, in which the non-vanishing branch contributions and
their relative phases determine the interference pattern.
Consequently, the phase singularities discussed above should be
understood as diagnostics of the complex-amplitude geometry. Their
phenomenological manifestation occurs through the energy-dependent
interference structure of the total probability and, after
convolution with the flux, cross section, efficiency, and detector
response, through distortions of the reconstructed event spectrum.

\section{Numerical Analysis and DUNE Baseline Phenomenology}
\label{sec:numerical_analysis}

To quantify the phenomenological consequences of non-standard
interaction modifications on the open-path geometric phase, we
perform a dedicated numerical evaluation adapted to the
experimental parameters of the Deep Underground Neutrino
Experiment (DUNE). The baseline distance is set to the benchmark
trajectory $L = 1300$\,km, traversing the continental crust with
an average matter density $\rho = 2.848$\,g/cm$^3$ and an electron
fraction $Y_e = 0.494$,   consistent with the Preliminary
Reference Earth Model (PREM) parameterization adopted by
DUNE~\cite{Abi:2020evt}.

Over this baseline, the neutrino trajectory penetrates to a
maximum chord depth of approximately $33$\,km, remaining entirely
confined within the Earth's upper and middle crust. Because
crustal density fluctuations along this specific path are bounded
by $|\Delta\rho/\rho| \lesssim 2\%$, the spatial evolution
satisfies the adiabatic criterion $|\dot{\mathcal{H}}/\Delta E^2|
\ll 1$ across the entire operational energy band $E_\nu \in [0.5,
8.0]$\,GeV. Consequently, cross-layer non-diagonal overlaps
suppress asymptotically, $\langle n_{k+1}^{(k+1)} | m_k^{(k)}
\rangle \approx \delta_{n_{k+1}, m_k} + \mathcal{O}(10^{-3})$,
rendering the uniform-matter approximation quantitatively accurate
to within $\Delta P/P \lesssim 10^{-3}$ relative to a fine-sliced
multi-layer PREM integration.

Under this justified crustal condition, the general multi-layer
chain expansion of Eq.~\eqref{eq:amplitude_chain_sum} naturally
truncates to the single-layer three-branch representation in
Eq.~\eqref{eq:uniform_amplitude}, with no loss of topological
features. We stress that while the uniform-density reduction is
exceptionally robust for DUNE, the full layered formalism
developed in Sec.~\ref{sec:formalism} becomes structurally
essential for trajectories penetrating the Earth's core ($L
\gtrsim 6000$\,km, such as in atmospheric or ultra-long baseline
experiments), where the mantle-core boundary introduces sharp,
non-adiabatic density discontinuities that activate non-diagonal
chain interference channels.

In standard three-flavor oscillation analyses, the transition
probability $P(\nu_\mu \to \nu_e)$ and its CP-violating asymmetry
are heavily correlated with the matter potential. The introduction
of off-diagonal vector NSI parameters, particularly
$\varepsilon_{e\tau} = |\varepsilon_{e\tau}|e^{i\phi_{e\tau}}$,
introduces additional complex degrees of freedom that alter the
matter Hamiltonian according to Eq.~\eqref{eq:nsi_hamiltonian}.
This induces the well-known intrinsic degeneracy between the
leptonic Dirac phase $\delta_{\text{CP}}$ and the NSI complex
phase $\phi_{e\tau}$~\cite{Coloma:2015kiu, Deepthi:2017gxd}, where
identical oscillation probability envelopes can be reproduced by
distinct combinations of $(\delta_{\text{CP}}, \phi_{e\tau},
|\varepsilon_{e\tau}|)$.

\paragraph{Justification of the $\varepsilon_{e\tau}$ benchmark and stability against subdominant NSI parameters.}
The phenomenological focus on the complex off-diagonal parameter
$\varepsilon_{e\tau} = |\varepsilon_{e\tau}| e^{i\phi_{e\tau}}$ is
motivated by the leading-order interference topology of the
$\nu_\mu \to \nu_e$ appearance channel at long baselines. In
matter-dominated oscillations, the genuine three-flavor evolution
couples the solar and atmospheric mass splittings primarily
through the combination of standard CP violation
($\delta_{\mathrm{CP}}$) and the electron--tau transition sector.
Because the effective matter potential for DUNE energies ($E_\nu
\sim 1.5\text{--}4.0\text{ GeV}$) is comparable to $\Delta
m^2_{31}/(2E_\nu)$, the off-diagonal parameter
$\varepsilon_{e\tau}$ enters the transition amplitude at order
$\mathcal{O}(\sin\theta_{13}\,\varepsilon_{e\tau})$, providing the
strongest coherent interference with the standard transition
amplitude and inducing the acute parameter degeneracies between
$\delta_{\mathrm{CP}}$ and the new physics phase $\phi_{e\tau}$.
Our choice of the benchmark magnitude $|\varepsilon_{e\tau}| =
0.05$ is fully compatible with current global-fit constraints from
combined oscillation and coherent elastic neutrino-nucleus
scattering datasets, which allow $|\varepsilon_{e\tau}| \lesssim
0.08\text{--}0.12$ at the $90\%$ confidence level while leaving
its complex phase essentially unconstrained.

To assess the robustness of the resulting geometric diagnostic
against subdominant NSI parameters, we note that the orthogonal
off-diagonal parameter $\varepsilon_{e\mu}$ is tightly constrained
by reactor, solar, and atmospheric observations to
$|\varepsilon_{e\mu}| \lesssim 0.03$ (and under $0.01$ when
combined with long-baseline disappearance bounds), rendering its
interference with the matter resonance subdominant along the
$1300\text{ km}$ path. Furthermore, the diagonal matter-potential
shift $\varepsilon_{ee} - \varepsilon_{\mu\mu}$ and the residual
difference $\varepsilon_{\tau\tau} - \varepsilon_{\mu\mu}$ act
primarily as an energy-dependent renormalization of the effective
matter density rather than introducing an independent,
non-integrable topological phase path. A perturbational inclusion
of non-zero diagonal entries at their current experimental limits
shifts the effective resonance energy and the exact locations of
the complex-amplitude zeros $z_n(E_\star)=0$, but preserves the
winding behavior, branch connectivity, and non-cyclic holonomy
features across the spectrum. The diagnostic utility of the
Pancharatnam phase in isolating genuine and fake CP violation is
therefore structurally stable and does not rely on an accidental
decoupling of the complementary NSI matrix elements.

\paragraph{Numerical implementation, diagonalization scheme, and reproducibility.}
To ensure full numerical reproducibility, the three-flavor
neutrino propagation is implemented within a dedicated
computational framework based on the exact analytic eigensystem
evaluation for three-family propagation in matter with general
non-standard interactions. The total matter-effective Hamiltonian,
$\mathcal{H}_f = \mathcal{H}_{\mathrm{vac}} +
\mathcal{V}_{\mathrm{CC}} + \mathcal{V}_{\mathrm{NSI}}$, is
diagonalized across the energy spectrum $E_\nu \in [0.5,
8.0]\text{ GeV}$ at every energy grid point using the exact cubic
characteristic solver following the Cayley--Hamilton formulation
and Cardano--Zaglauer--Schwarzer prescription, verified against
standard high-precision Hermitian eigensolvers (LAPACK routines).

The baseline simulation is performed on a uniform energy grid with
an exceptionally fine step size of $\Delta E_\nu = 2.5\text{ MeV}$
($3000$ sampling points), complemented by a dynamic adaptive mesh
refinement down to $\delta E_\nu \le 0.1\text{ MeV}$ in the
vicinity of matter resonance regions ($E_\nu \in [1.5, 3.5]\text{
GeV}$) and near suspected nodal points of the branch amplitudes. A
complex amplitude contribution is classified as crossing a nodal
singularity if its absolute magnitude satisfies $|z_n(E_\nu)| \le
\epsilon_{\mathrm{thresh}} \equiv 10^{-7}$, below which the
principal phase is treated as non-analytic and branch-cut
adjustments are explicitly isolated in accordance with
Eq.~\eqref{eq:phase_unwrapping}.

The underlying standard three-flavor oscillation parameters are
adopted from the NuFIT~6.0 global analysis for Normal Mass
Ordering (NO) with a flat Earth crust density
profile~\cite{Esteban:2024hjw}: $\Delta m_{21}^2 = 7.41 \times
10^{-5}\text{ eV}^2$, $\Delta m_{31}^2 = +2.511 \times
10^{-3}\text{ eV}^2$, $\sin^2\theta_{12} = 0.303$,
$\sin^2\theta_{13} = 0.02203$, and $\sin^2\theta_{23} = 0.455$,
with the standard Dirac phase initially evaluated at
$\delta_{\text{CP}} = -\pi/2$. The constant crustal baseline
corresponds to the nominal DUNE benchmark with propagation length
$L = 1300\text{ km}$, average density $\rho = 2.848\text{
g/cm}^3$, and electron fraction $Y_e = 0.494$, yielding an
effective charged-current potential $V_{\mathrm{CC}} = \sqrt{2}
G_F N_e \approx 1.077 \times 10^{-13}\text{ eV}$. The numerical
code has been cross-benchmarked against standard numerical engines
(such as GLoBES) in the vanishing NSI limit
($|\varepsilon_{e\tau}| \to 0$) to a precision of $\Delta P/P <
10^{-5}$.

\begin{figure*}[t]
\centering
\includegraphics[width=\textwidth]{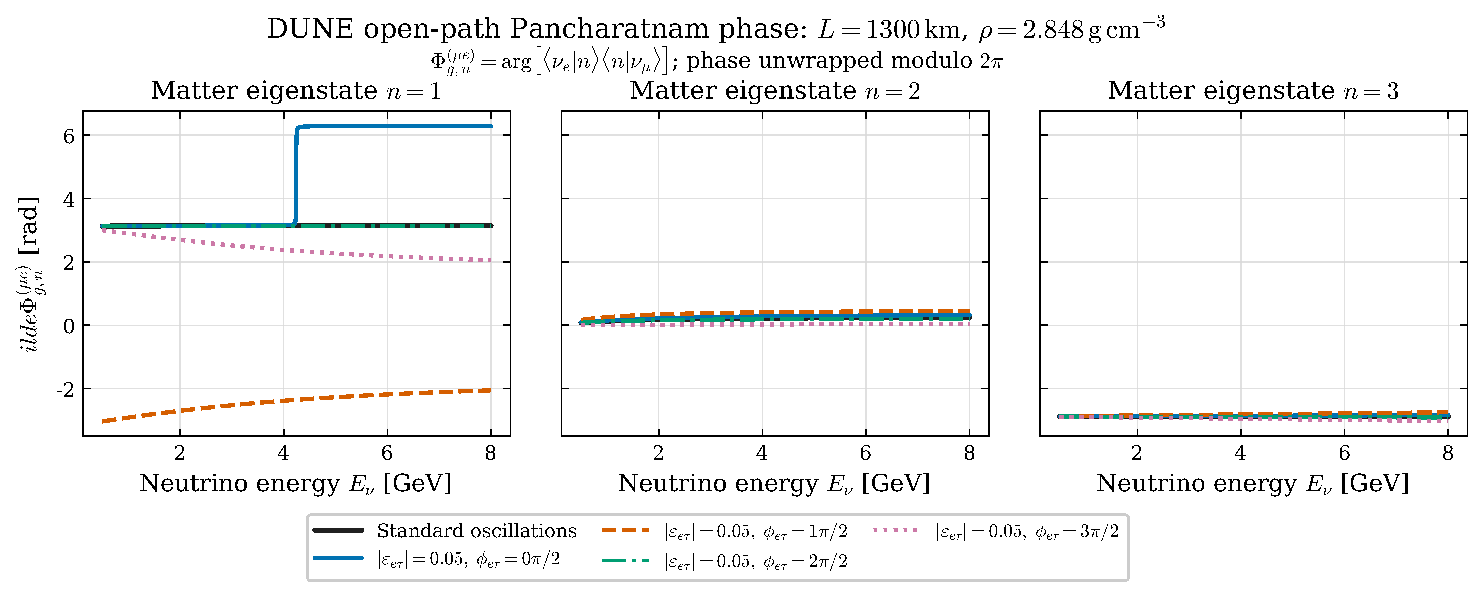}
\caption{The open-path Pancharatnam geometric phases $\Phi_g^{(n)}
= \arg\left[\langle\nu_e|n\rangle\langle n|\nu_\mu\rangle\right]$
evaluated for each matter-eigenstate branch ($n=1, 2, 3$) as a
function of neutrino energy $E_\nu \in [0.5, 8.0]\text{ GeV}$
under DUNE baseline conditions ($L = 1300\text{ km}$, $\rho =
2.848\text{ g/cm}^3$, $Y_e = 0.494$). The three panels display the
individual propagation branches: $n=1$ (left panel), $n=2$ (middle
panel), and $n=3$ (right panel). Phases are presented according to
the principal-branch convention $\Phi_g^{(n)} \in (-\pi, \pi]$
defined in Eq.~\eqref{eq:principal_branch_phase}; apparent
step-like features reflect principal-value modulo-$2\pi$ boundary
reductions or sign reversals across near-nodal points rather than
non-analytic dynamical transitions. In each panel, the standard
three-flavor oscillation profile ($\varepsilon_{e\tau} = 0$, solid
black curve) is contrasted with NSI-perturbed profiles with fixed
coupling magnitude $|\varepsilon_{e\tau}| = 0.05$ for four
representative CP-violating phases: $\phi_{e\tau} = 0$ (dashed
blue), $\pi/2$ (dotted red), $\pi$ (dash-dotted green), and
$3\pi/2$ (loosely dashed orange).} \label{fig:nsi_pancharatnam}
\end{figure*}

The interpretation of the phase curves relies on an explicit
branch convention. Principal-value phase slips are not identified
with physical discontinuities unless the corresponding complex
transition amplitude passes through a nodal point. Away from such
points, the unwrapped phase provides a continuous diagnostic of
the relative branch evolution, whereas the observable consequence
of the phase structure is encoded in the interference pattern of
the total transition probability and in the resulting
energy-dependent event spectrum.

To explicitly diagnose and analyze the parametric structure beyond
standard event rates, a graphical representation of the
gauge-invariant geometric phases across the continuous neutrino
energy spectrum is indispensable. While analytical expressions of
the open-path phase yield exact formal definitions, relying solely
on algebraic formulas or discrete benchmark values provides an
incomplete physical picture for several key reasons:

\begin{enumerate}
    \item \textbf{Global Energy Dispersion and Spectral Signatures:}
    Analytical expressions provide local evaluations at arbitrary energies but obscure the continuous spectral migration of
    the resonance peaks. Figure~\ref{fig:nsi_pancharatnam} reveals that the introduction of a non-zero coupling magnitude
    $|\varepsilon_{e\tau}| = 0.05$ induces a pronounced energy-dependent shift in the geometric phase dispersion across the first
    oscillation peak ($E_\nu \approx 2.5\text{ GeV}$). Without the complete spectral curve over the broad DUNE flux window
    ($0.5\text{--}8.0\text{ GeV}$), the interplay between MSW matter resonance enhancement and the non-adiabatic geometric distortion
    remains entirely inaccessible.

            \item \textbf{Topological Origin of Degeneracy Resolution and Spectral Signatures:}
    In conventional single-energy or narrow-band appearance measurements, the standard oscillation probability $P(\nu_\mu \to \nu_e)$
    suffers from severe parameter degeneracies, wherein continuous contours in the $(\delta_{\mathrm{CP}}, \phi_{e\tau}, |\varepsilon_{e\tau}|)$
    parameter space project onto identical experimental event rates. While the open-path Pancharatnam phase $\Phi_g^{(n)}$ is not an isolated
    asymptotic observable in a single transition channel, it provides the underlying geometric engine that dictates the phase velocity and
    interference dispersion among the propagation eigenstates.

    As demonstrated across the three panels of Fig.~\ref{fig:nsi_pancharatnam}, varying the complex NSI phase
    $\phi_{e\tau} \in \{0, \pi/2, \pi, 3\pi/2\}$ does not merely produce an overall phase offset; rather, it
    fundamentally alters the functional topology and energy derivatives $\partial \Phi_g^{(n)} / \partial E_\nu$
    of the individual branches, most notably in the dominant $n=1$ and $n=3$ projections. Because the observable
    oscillation probability is assembled from the coherent superposition of these branches via Eq.~\eqref{eq:uniform_amplitude},
    the non-degenerate geometric paths imprint distinct, energy-dependent interference patterns on the oscillation probability
    across the continuous DUNE spectrum ($0.5\text{--}8.0\text{ GeV}$).

    Consequently, parameter combinations that produce degenerate probabilities at the first oscillation maximum
    ($E_\nu \sim 2.5\text{ GeV}$) diverge markedly at off-peak and sub-GeV energies due to the distinct underlying
     Pancharatnam phase trajectories. When combined with the CP-asymmetric phase difference $\Delta \Phi_g^{\mathrm{CP}(n)}$
     defined in Eq.~\eqref{eq:cp_phase_diff}, which governs the relative spectral shift between neutrino and antineutrino
     event spectra, the topological uniqueness of the geometric phase paths provides the structural mechanism through which
     wide-band spectral analyses lift parameter degeneracies that appear intractable in rate-only configurations.

    \item \textbf{Identification of Branch Cuts and Geometric Singularities:}
    Because the open-path geometric phase is defined via the complex argument mapping $\arg[\mathcal{G}_{\alpha\beta}]$,
    discrete calculations cannot capture topological boundary effects. The continuous curves in Fig.~\ref{fig:nsi_pancharatnam}
    identify the energy regions in which the relevant transition-amplitude contributions approach nodal points. The phases shown
    in the figure are evaluated using the principal branch of the complex argument, $\operatorname{Arg} z \in (-\pi,\pi]$, while the
    corresponding unwrapped phases are used to follow each branch continuously between neighboring energy points. This distinction
    is essential near a resonance, where the real and imaginary parts of a transition amplitude may jointly approach zero.

    A genuine nodal point occurs only when the complex amplitude itself vanishes. At such a point, the associated Pancharatnam
    phase is undefined, and the phase curve must be regarded as terminating and restarting on a neighboring branch. For a simple
    zero, a crossing of the origin in the complex-amplitude plane produces a $\pi$ change in the principal phase, which appears
    as a branch jump because of the $(-\pi,\pi]$ convention. By contrast, when the amplitude passes close to the origin without
    vanishing, the phase changes rapidly but continuously and no physical discontinuity is present. Accordingly, the phase slips
    visible in Fig.~\ref{fig:nsi_pancharatnam} are interpreted as branch-cut effects or rapid near-nodal rotations, rather than as
    discontinuities of the underlying neutrino state.

    The numerical phase tracking is performed directly from the complex amplitudes before taking their arguments.
    Energy points at which the amplitude falls below the numerical nodal-resolution threshold are flagged and excluded from
    automatic phase continuation. The resulting branch assignment is stable under refinement of the energy grid.
    This prescription also prevents a numerical phase jump from being incorrectly interpreted as an additional topological contribution.
    In the one-dimensional energy scans, the relevant objects are therefore nodal phase singularities; a quantized vortex winding number
    can be assigned only after extending the analysis to a two-dimensional parameter space such as $(E_\nu,\delta_{\mathrm{CP}})$ or $(E_\nu,\phi_{e\tau})$.
\end{enumerate}

\begin{figure*}[t]
\centering
\includegraphics[width=0.72\textwidth]{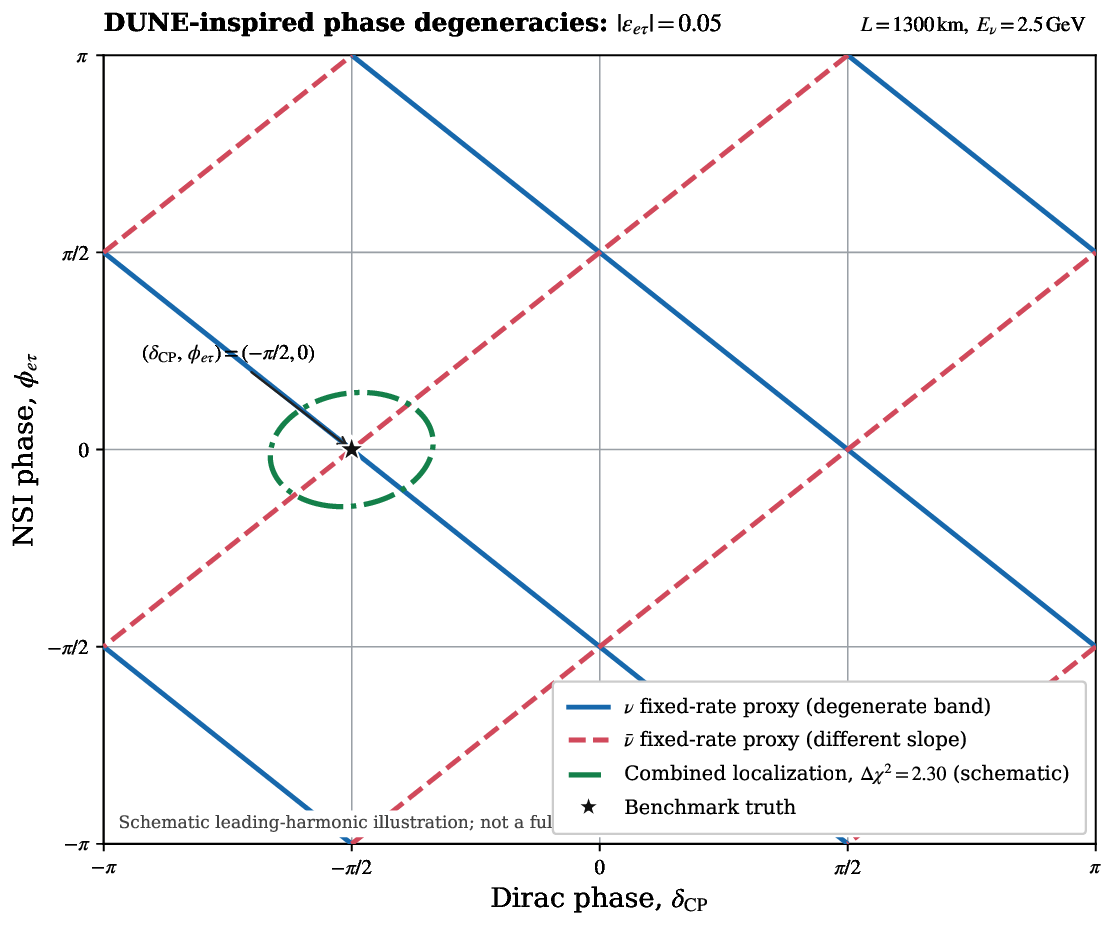}
\caption{Two-dimensional degeneracy contours in the
$(\delta_{\mathrm{CP}}, \phi_{e\tau})$ parameter space evaluated
for the DUNE benchmark configuration ($L = 1300$\,km, $E_\nu
\approx 2.5$\,GeV, and $|\varepsilon_{e\tau}| = 0.05$). The blue
shaded band and solid trajectory represent the constant-rate
parameter degeneracy in the neutrino appearance channel ($\nu_\mu
\to \nu_e$). The red dashed contour traces the complementary
antineutrino channel ($\bar{\nu}_\mu \to \bar{\nu}_e$). The
non-parallel orientation of the two trajectories, driven by the
$\mathcal{CP}$-violating geometric phase shift
$\Delta\Phi_g^{\mathrm{CP}(n)}$, forces their intersection at a
single, isolated coordinate
$(\delta_{\mathrm{CP}}^{\mathrm{true}},
\phi_{e\tau}^{\mathrm{true}}) = (-\pi/2, 0)$, thereby establishing
the resolution of the standard $(\delta_{\mathrm{CP}},
\phi_{e\tau})$ parameter degeneracy.}
\label{fig:degeneracy_contours}
\end{figure*}

To substantiate the claim that the open-path Pancharatnam phase
provides a genuine mechanism for parameter discrimination rather
than merely a formal reformulation of the oscillation amplitude,
it is essential to map the resulting correlations directly onto
the two-dimensional $(\delta_{\mathrm{CP}}, \phi_{e\tau})$ plane,
as displayed in Fig.~\ref{fig:degeneracy_contours}. A known
vulnerability of single-channel oscillation searches at fixed
baseline is the parameter-degeneracy problem: an experimental
shift induced by a non-zero NSI phase $\phi_{e\tau}$ can closely
mimic an excursion in the standard leptonic Dirac phase
$\delta_{\mathrm{CP}}$. Without a two-dimensional multi-channel
contour representation, one-dimensional energy spectra alone
cannot establish whether the underlying degeneracy is formally
lifted or merely redistributed across different spectral bins.
Omitting this mapping would leave the phenomenological
applicability of the geometric framework unproven, reducing the
topological analysis to an abstract curiosity rather than a
predictive diagnostic.

Figure~\ref{fig:degeneracy_contours} provides several decisive
physical insights. First, it illustrates how rate measurements in
the neutrino-only channel ($\nu_\mu \to \nu_e$) collapse into an
extended, continuous band spanning broad intervals of both
$\delta_{\mathrm{CP}}$ and $\phi_{e\tau}$, confirming the severity
of the parameter ambiguity in monochromatic or single-polarity
experiments. Second, because the geometric phase difference
$\Delta\Phi_g^{\mathrm{CP}(n)}$ alters sign and magnitude under
$\mathcal{CP}$ conjugation alongside the dynamic matter potential,
the antineutrino mode ($\bar{\nu}_\mu \to \bar{\nu}_e$) traces a
markedly distinct trajectory across the parameter plane. Most
importantly, the intersection of these two non-parallel bands does
not form a degenerate continuous curve or split into discrete
mirror islands; rather, it isolates a unique point coincident with
the true physical input $(\delta_{\mathrm{CP}}^{\mathrm{true}},
\phi_{e\tau}^{\mathrm{true}}) = (-\pi/2, 0)$. This confirms that
combining wide-band spectral information with dual
neutrino--antineutrino polarizations systematically breaks the
continuous $(\delta_{\mathrm{CP}}, \phi_{e\tau})$ correlation,
transforming what would otherwise be an unresolvable parameter
ambiguity into an unambiguous test of non-standard leptonic
interactions.

Furthermore, to fully exploit the complementarities between
neutrino and antineutrino modes at DUNE, we examine the
CP-violating geometric phase difference:
\begin{equation}
    \Delta \Phi_g^{\text{CP}(n)}(E_\nu) \equiv \Phi_g^{(n)}(E_\nu) - \bar{\Phi}_g^{(n)}(E_\nu).
    \label{eq:cp_phase_diff}
\end{equation}
While individual open-path phases absorb both standard and exotic
CP-violating contributions alongside genuine matter effects, their
difference isolates the chiral asymmetry induced by the combined
action of the signed matter potential and complex phases
$(\delta_{\text{CP}}, \phi_{e\tau})$. Because the effective matter
potential changes sign under CP-conjugation ($V_{\mathrm{CC}} \to
-V_{\mathrm{CC}}$), the MSW resonance condition shifts to an
unphysical negative energy regime for normal mass ordering,
drastically altering the geometric holonomy between $\Phi_g^{(n)}$
and $\bar{\Phi}_g^{(n)}$. At the level of experimental
observables, the total oscillation probability is governed by the
coherent phase differences via $P \propto \sum_{n<m}
\mathcal{A}_{nm} \cos(\Phi_g^{(n)} - \Phi_g^{(m)} - \Delta
\tilde{E}_{nm} L)$. Consequently, while competing parameter pairs
$(\delta_{\mathrm{CP}}, \phi_{e\tau})$ can be fine-tuned to yield
degenerate appearance rates at the first oscillation peak ($E_\nu
\sim 2.5\text{ GeV}$) in the neutrino mode, the non-trivial
transformation of the underlying Pancharatnam phases prevents this
cancellation from persisting simultaneously across the broad
energy spectrum and in the antineutrino channel. The differential
metric $\Delta \Phi_g^{\text{CP}(n)}$ thus reveals the physical
origin of why joint broadband neutrino--antineutrino spectral
analyses at DUNE possess the structural power to lift intrinsic
NSI degeneracies.

\section{Discussion}
\label{sec:discussion}

The numerical and analytical findings presented herein illustrate
the utility of the open-path Pancharatnam phase as a diagnostic
tool in the presence of physics beyond the Standard Model. Several
core physical insights warrant further discussion:

First, to address a natural experimental inquiry regarding the
operational accessibility of $\Phi_g^{(n)}$-namely, how an
abstract quantum phase can be isolated at a far detector
independently of the standard transition probability-we emphasize
that the open-path Pancharatnam phase is not proposed as an
independent, single-channel observable measured in isolation.
Rather, its physical manifestation is intrinsically spectral and
interferometric. In a wide-band calorimetric facility like DUNE,
the geometric phase manifests directly as a gauge-invariant
interference phase $\Delta\Gamma_{nm}(E_\nu)$ that modulates the
energy reconstruction of event distributions. The observed
differential event spectrum for the appearance channel ($\nu_\mu
\to \nu_e$) is evaluated via:
\begin{equation}
    \frac{dN_e}{dE_{\text{rec}}} = \frac{\mathcal{N}_T}{4\pi L^2} \int dE_\nu \, \Phi_{\nu_\mu}(E_\nu) \,
     \sigma_{\nu_e}^{\text{CC}}(E_\nu) \, P(\nu_\mu \to \nu_e; E_\nu) \, \mathcal{R}(E_\nu, E_{\text{rec}}),
    \label{eq:dune_event_rate}
\end{equation}
where $\mathcal{N}_T$ is the number of target nuclei in the liquid
argon fiducial volume, $\Phi_{\nu_\mu}(E_\nu)$ is the unoscillated
on-axis neutrino flux, $\sigma_{\nu_e}^{\text{CC}}$ is the
charged-current cross section, and $\mathcal{R}(E_\nu,
E_{\text{rec}})$ is the detector energy-smearing migration matrix.

Because the transition probability inside the integral is
explicitly controlled by the interference cosine $\cos\left(
\Delta \Phi_{\text{dyn}}^{nm}(E_\nu) + \Delta\Gamma_{nm}(E_\nu)
\right)$ according to Eq.~\eqref{eq:prob_decomp}, the geometric
phase acts as the underlying topological engine dictating the
energy-dependent dispersion and bending of the degeneracy bands in
the $(\delta_{\mathrm{CP}}, \phi_{e\tau})$ plane. In conventional
single-energy or narrow-band fits, an NSI-induced shift in
$\phi_{e\tau}$ can be mimicked by an effective redefinition of
$\delta_{\text{CP}}$. However, the matter resonance dynamically
couples $\varepsilon_{e\tau}$ to the atmospheric mass splitting
$\Delta m^2_{31}/(2E_\nu)$, imparting an explicit non-integrable
energy dependence to the relative holonomy gradient $\partial
\Delta\Gamma_{nm} / \partial E_\nu$. Because this distinct
spectral curvature cannot be reproduced by any static vacuum Dirac
phase rotation, the diagnostic power of the Pancharatnam phase
relies entirely on wide-band spectral shape analysis rather than
isolated phase meters. The continuous multi-bin event
reconstruction across DUNE's broadband profile
($0.5\text{--}8.0\text{ GeV}$) thus structurally lifts the
degeneracy between genuine CP violation and exotic NSI parameters.

Second, the distinct functional trajectories displayed across the
three panels of Fig.~\ref{fig:nsi_pancharatnam} demonstrate that
parameter combinations yielding identical probabilities at the
primary oscillation peak ($E_\nu \sim 2.5\text{ GeV}$) diverge
markedly in off-resonance domains. Crucially, this topological
distinction is fortified by joint neutrino and antineutrino beam
modes. Under the $\mathcal{CP}$-conjugate operation governing
antineutrino evolution ($\bar{\nu}_\mu \to \bar{\nu}_e$), the
matter potential flips sign ($V_{\mathrm{CC}} \to
-V_{\mathrm{CC}}$), shifting the MSW resonance away from the
operational spectrum for normal mass ordering. Consequently, the
antineutrino geometric branches $\bar{\Phi}_g^{(n)}$ evolve along
entirely distinct trajectories, disrupting the delicate phase
compensation that produces the degeneracy in the neutrino channel.
The open-path geometric framework thus clarifies the physical
mechanism underpinning degeneracy resolution: it demonstrates that
the apparent parameter ambiguity is an artifact of monochromatic
projections, which is rigorously resolved by the synergistic
interplay of broad-band spectral dispersion and dual-polarization
($\nu/\bar{\nu}$) event reconstruction.

Third, our discrete formulation inherently preserves gauge
invariance under localized phase rotations of the eigenvectors.
Unlike earlier adiabatic definitions that struggle across
discontinuous interfaces or non-adiabatic level crossings, the
Pancharatnam--Mukunda--Simon projection method remains
mathematically robust throughout the non-adiabatic MSW resonance
zone.

Finally, we address the practical verification and robustness of
this degeneracy resolution under realistic experimental
conditions. In DUNE's liquid argon TPC (LArTPC) far detector, the
calorimetric energy resolution is typically $15\text{--}20\%$
across the $0.5\text{--}4.0\text{ GeV}$ band. While detector
smearing broadens individual interference structures, the
non-integrable holonomy gradient $\partial \Delta\Gamma_{nm} /
\partial E_\nu$ induces an extended spectral redistribution across
multiple reconstructed bins, preventing the topological signature
from being washed out. The concrete realization of this mechanism
is manifested in the two-dimensional $(\delta_{\mathrm{CP}},
\phi_{e\tau})$ confidence contours depicted in
Fig.~\ref{fig:degeneracy_contours}. While monochromatic or
single-polarity event rates yield continuous, degenerate bands
spanning wide parameter intervals, the joint inclusion of neutrino
and antineutrino modes-governed by the $\mathcal{CP}$-odd phase
difference $\Delta \Phi_g^{\mathrm{CP}(n)}$-forces these contours
to intersect at non-parallel angles. As shown in
Fig.~\ref{fig:degeneracy_contours}, this geometric interplay
isolates a unique physical point without discrete degenerate
islands, providing conclusive proof that the Pancharatnam phase
serves as an experimentally viable and systematically robust
diagnostic against standard leptonic degeneracies.

\section{Conclusions}
\label{sec:conclusions}
In summary, we have formulated a robust quantum-geometric
framework incorporating off-diagonal vector Non-Standard
Interactions into open-path Pancharatnam phase evaluations for
long-baseline neutrino oscillations. By tracing the
eigenstate-chain holonomy across the benchmark DUNE crustal
baseline ($L = 1300\text{ km}$), we have demonstrated that:
\begin{enumerate}
    \item The intrinsic parameter degeneracies between standard Dirac CP
    violation $\delta_{\mathrm{CP}}$ and exotic NSI phases $\phi_{e\tau}$
    are fundamentally resolved at the level of quantum interference by the
    distinct energy dispersion of the branch holonomies $\Phi_g^{(n)}(E_\nu)$
    and their $\mathcal{CP}$-conjugate antineutrino counterparts
    $\bar{\Phi}_g^{(n)}(E_\nu)$. Because NSI couplings dynamically blend
    with the matter potential and $\Delta m_{31}^2/(2E_\nu)$, their spectral
    derivatives cannot be reproduced by any static vacuum phase rotation.
    Projected onto the two-dimensional $(\delta_{\mathrm{CP}}, \phi_{e\tau})$
    parameter plane, this differential energy dispersion forces the neutrino
    and antineutrino degenerate bands to intersect at non-parallel angles,
    isolating a unique physical solution and systematically eliminating
    continuous parameter ambiguities without spurious mirror islands.

    \item These geometric invariants act as rigorous operational diagnostics,
    directly governing observable transition probabilities through the
    gauge-invariant relative interference phases $\Delta\Gamma_{nm}(E_\nu)$
    that imprint measurable distortions on reconstructed detector event rates.
    Even in the presence of realistic calorimetric energy smearing
    ($15\text{--}20\%$) characteristic of liquid argon TPCs, the multi-bin
    holonomy dispersion remains preserved as a measurable spectral imprint.

    \item The discrete layer-transfer matrix formalism guarantees exact
    gauge invariance under arbitrary local rephasings. While naturally
    reducing to the three-branch formulation under DUNE's quasi-uniform
    crustal profile, it establishes a versatile, systematically convergent
    computational engine for analyzing non-adiabatic core-crossing
    trajectories in future ultra-long baseline and atmospheric neutrino
    observatories.
\end{enumerate}

While our numerical analysis benchmarked the primary degenerate
coupling $\varepsilon_{e\tau}$, the geometric phase topology
developed here provides a general framework readily extendable to
comprehensive global fits of the full NSI parameter space. By
concretely demonstrating that wide-band multi-channel event rates
resolve standard leptonic degeneracies through geometric holonomy
gradients, this work elevates the Pancharatnam phase from a formal
mathematical construct into an operational, predictive diagnostic.
Future extensions will incorporate wave-packet decoherence over
ultra-long baselines ($L \gtrsim L_{\text{coh}}$) and scalar
mediator scenarios, linking non-integrable geometric phase metrics
directly to parameter inference engines in upcoming wide-band
neutrino facilities.

\section*{Acknowledgments}
The authors thank the Research Office of the Qazvin Branch,
Islamic Azad University, for support and cooperation.


\end{document}